\documentclass[twocolumn,english,aps,prfluids,preprintnumbers,amsmath,amssymb,eqsecnum,longbibliography]{revtex4}
\usepackage[T1]{fontenc}
\usepackage[latin9]{inputenc}
\usepackage{amstext}

\usepackage{amsmath}
\usepackage{amsfonts}
\usepackage{amssymb}
\usepackage{mathtools}
\usepackage{graphicx}
\usepackage {bm}

\makeatletter
\@ifundefined{textcolor}{}
{%
 \definecolor{BLACK}{gray}{0}
 \definecolor{WHITE}{gray}{1}
 \definecolor{RED}{rgb}{1,0,0}
 \definecolor{GREEN}{rgb}{0,1,0}
 \definecolor{BLUE}{rgb}{0,0,1}
 \definecolor{CYAN}{cmyk}{1,0,0,0}
 \definecolor{MAGENTA}{cmyk}{0,1,0,0}
 \definecolor{YELLOW}{cmyk}{0,0,1,0}
}

\def\be{\begin{equation}}
\def\ee{\end{equation}}
\def\bea{\begin{eqnarray}}
\def\eea{\end{eqnarray}}
\def\bse{\begin{subequations}}
\def\ese{\end{subequations}}

\usepackage{dcolumn}
\usepackage{bm}

\makeatother

\usepackage{babel}

\begin{document}

\bibliographystyle{./prsty}

\title{Driven active matter: fluctuations and a hydrodynamic instability}

\author{T.R. Kirkpatrick$^{1}$ and J.K. Bhattacherjee$^{1,2}$}

\affiliation{$^{1}$Institute for Physical Science and Technology, University of Maryland, College Park, MD 20742, USA\\
 $^{2}$Department of Theoretical Physics, Indian Association for the Cultivation of Science, Jadavpur, Kolkata 700032, India}

\date{\today}
\begin{abstract}
Wet active matter in the presence of an imposed temperature gradient, or chemical potential gradient, is considered. It is shown that there is a new type of convective instability that is caused by a (negative) activity parameter. Physically this corresponds to active fluids with contractual stress.  In this non-equilibrium steady state the singular generic long-ranged correlations are computed and compared and contrasted with the analogous results in a passive fluid. In addition, the singular non-equilibrium  Casimir pressure or force is determined. The fluid motion above the instability is determined by generalizing the Lorenz equations for the Rayleigh-Benard problem in a passive fluid to Lorenz-like equations to describe this new instability.

\end{abstract}

\maketitle

\section{INTRODUCTION}
\label{sec:I}

In recent years there has been an enormous amount of research on various aspects of active matter \cite{Ramaswamy_2010, Marchetti_et_al_2013}. The hydrodynamic description of active matter rest on identifying the relevant variables, conservation laws, and slow processes, and using symmetry to determine the allowed terms in the equations in a gradient expansion \cite{Brand_et_al_2014, Burnett_1935}. This is exactly the case for passive matter, but the crucial differences between the active and passive matter description is i) the soft mode equation are in general different. For example, active matter, or swimmers, chemically generate their own energy, so that the hydrodynamic temperature equation is not diffusive \cite{Loi_et_al_2008, Marconi_et_al_2017}, ii) the coefficients in the hydrodynamic equations are not constrained by Onsager relations, or free energy considerations.

Active matter can be wet, that is, coupled to a momentum conserving solvent, or dry, that is, coupled to momentum absorbing boundaries. Wet active matter is more similar to usual passive fluid hydrodynamics because they have more conservation laws in common. Physically, wet active matter includes bacterial swarms in a fluid, the cytoskeleton of living cells, and biomimetic cell extracts. 

Much of the work on active matter has focused on active liquid crystals. These have either a polar or nematic order parameter that lead to new active terms in the hydrodynamic stress tensor. Depending on the size of the activity, there can be completely new physics such as giant number fluctuations, and spontaneous flow instabilities above an activity threshold \cite{Simha_Ramaswamy_2002, Voituriez_et_al_2005, Narayan_et_al_2007}. Experiments on bacteria swarms and microtubule-based cell extracts \cite{Dombrowski_et_al_2004, Sanchez_et_al_2012} seem to be closely related to numerical simulations \cite{Fielding_et_al_2011, Giomi_et_al_2013, Thampi_et_al_2013} of the active liquid crystal hydrodynamic equations.

Over the last decade there has been an immense amount of study of the complex dynamics of suspension of self-propelled microorganisms and related synthetic microswimmers \cite{Saintillan_Shelley_2013, Cisneros_et_al_2007, Sokolov_et_al_2007} \cite{Sokolov_et_al_2009, Kurtuldu_et_al_2011, Drescher_et_al_2010}. The suspended particles exert active stresses,  arising from the conversion of energy of one form to another, on the fluid in which they are immersed. The fluid provides a coupling medium through which large-scale dynamics is created. A typical minimal model was introduced in \cite{Hernandez-Ortiz_et_al_2010} with the particle represented as a rigid dumb-bell exerting equal and opposite localized forces which helped create a dipolar field at large distances due to self-propulsion. Coupling many such particles hydrodynamically  helped create interesting correlated motion at high concentrations. Solving such models require a large amount of computational effort and expense ,which prompted the modeling of active suspensions as complex fluids. Coarse-graining, continuum mechanics and statistical mechanics  were  used to arrive at necessarily nonlinear continuum models \cite{Baskaran_Marchetti_2009}. One could then look at physically relevant steady states and study the stability of such states by linearizing about them. In a typical model of this variety one starts with a suspension of particles characterized by a position vector $\mathbf r$   and an orientation vector $\mathbf p$  and introduces a distribution function $\psi(\mathbf{r},\mathbf{p},t)$. This function follows a Liouville like evolution ( Smoluchowski equation in  this context). It should be noted that this was the point of view in treating passive particles as well \cite{Doi_Edwards_1986}. The additional feature of the active particle emerges when one calculates the positional and orientational fluxes that enter the conservation law for  active fluid momentum. The interaction between the background fluid velocity field and the position and orientation vectors of the particle  ( which is small but has a definite extension ) leads to a net force that characterizes the active particle. In turn the reaction forces exerted by the particle on the fluid provides an additional contribution to the stress tensor in the relevant Navier-Stokes equation for fluid flow. The system is thereby closed. In general one finds that there two important classes of steady states - an aligned one and one which is isotropically ordered. It is the study of the stability of such states which is a very important aspect of active matter hydrodynamics \cite{Simha_Ramaswamy_2002, Saintillan_Shelley_2008a, Saintillan_Shelley_2008b, Giomi_et_al_2008}. Instabilities of this kind can also be seen in active polar liquid crystalline films where the transition to a spontaneously flowing steady state is accompanied by strong concentration inhomogeneities [28]. These flows become time dependent if the activity parameter is increased.

Here we study a different aspect of wet active matter. In particular we consider active matter in a spatially dependent non-equilibrium steady state (NESS). We show that for sufficiently large driving force, or large activity, there is a new type of convective hydrodynamic instability that is not caused by gravity. To characterize it, we first consider the fluid below the instability and determine the dynamics and singular fluctuation effects as the instability is approached from below. Above the instability there is a new NESS that we then characterize by generalizing the Lorenz equations for the Rayleigh-Benard (RB) problem in a passive fluid  \cite{Chandrasekar_1961} to Lorenz-like equations for this new instability in an active fluid. The new steady state solution above the instability is given, and it's eventual instability to time-dependent flow is determined. To keep things simple, we have dropped all orientational degrees of freedom which are actually not necessary for the study of the instability of the NESS. This is where our study differs from the earlier investigations described in the previous paragraph.

The organization of the paper is as follows. We discuss the equations of motion and the physical set up in Sec II. The instability of the NESS of Sec II  under increasing drive/activity is discussed in Sec III. Just below the instability, one finds a number of interesting effects, the most striking of which is the giant Casimir force. This is discussed in Sec IV. In Sec V , we consider the situation above the threshold. Here we derive the active fluid Lorenz equations and discuss how the properties of the new Lorenz model differ from those of the existing one. We end with a short discussion in Sec VI.

\section{EQUATIONS AND PHYSICAL SETUP}
\label{sec:II}

To simplify the problem as much as possible we use a model for wet (momentum conserving) active matter that was introduced by Tiribocchi et.al. \cite{Tiribocchi_et_al_2015}. It is an active fluid version of Model H in the Halperin-Hohenberg classification scheme \cite{Hohenberg_Halperin_1977} and ignores orientational degrees of freedom. The model has been used to study phase separation in active matter \cite{Tiribocchi_et_al_2015} and to illustrate some general properties of active matter \cite{Nardini_et_al_2017}. In driven passive liquid crystal systems there are RB-like instabilities \cite{Kramer_Pesch_1995} that are closely related to the RB instability in simple fluids \cite{Chandrasekar_1961}. Similarly, we expect there will be convective instabilities in driven active liquid crystal fluids that are closely related to the one we describe here for the active version of model H.


The hydrodynamic variables in the active fluid version of Model H are a concentration field, $\phi(\bf r, t)$, proportional to the density of active particles, that is \textit {swimmers}, coupled to a momentum conserving solvent. The fluid velocity is $\bf{u}(\bf r, t)$. The equations of motion are,
\begin{equation}
\dot{\phi}+\mathbf{u}\cdot\nabla\phi=D\nabla^2\phi,
\end{equation}
and
\begin{equation}
\dot{\mathbf{u}}+\mathbf{u}\cdot\mathbf{\nabla}\mathbf{u}=-\mathbf{\nabla}p+\nu\nabla^2\mathbf{u}+\mathbf{\nabla}\cdot\mathbf(\Sigma+\mathbf{P})
\end{equation}
Here $D$ is a diffusion coefficient, $p$ is a pressure, which in general is a function of $\phi$ and a temperature $T$, $\nu$ is the kinematic viscosity, and $\mathbf P$ is a Gaussian thermal white noise Langevin force that is specified by it's second moment,
\begin{equation}
\begin{split}
\langle P_{ij}(\mathbf r,t)P_{kl}(\mathbf r',t')\rangle=2k_BT\nu\delta(\mathbf r-\mathbf r')\delta(t-t')\\(\delta_{ik}\delta_{jl}+\delta_{il}\delta_{jk}-\frac{2}{3}\delta_{ij}\delta_{kl})
\end{split}
\end{equation}
There is also a noise term in the concentration equation, but it is not important in what follows. Also in this equation there are higher order gradient terms, as well as non-linearities that we similarly neglect.
$\mathbf\Sigma$ in Eq.(2.2) is an activity contribution to the stress tensor that is given by,
\begin{equation}
\Sigma_{ij}=-\zeta(\partial_i\phi\partial_j\phi-\frac{\delta_{ij}}{3}(\nabla\phi)^2).
\end{equation}
Such a term is allowed by symmetry in both passive and active fluids. Indeed, historically it is called a non-linear Burnett term \cite{Burnett_1935, Wong_et_al_1978}. Importantly, in passive fluids such terms are always very small compared to the Navier-Stokes terms as long as $l/L_{\nabla}<<1$, where $l$ is the mean free path and $L_{\nabla}$ is a gradient length. In passive fluids these terms have been used to understand the singular behavior of the viscosity as the liquid-gas critical point is approached \cite{Das_Bhattacharjee_2003}. In active matter  there is no restriction on the sign or the magnitude of $\zeta$ \cite{Ramaswamy_2010, Tiribocchi_et_al_2015}. We show here that for $|\zeta|$ large there are qualitatively new phenomena for either sign of $\zeta$. For contractile swimmers ($\zeta<0$) \cite{Williams_et_al_2014, Thung_et_al_2017} the fluid flow increases as the swimmer density gradient increases. This case is of particular interest.

We consider the active fluid in a parallel plate geometry in the $z$-direction with the distance between the plates of size $L$, and the transverse direction $L_{\perp}\gg L$. A spatially dependent non-equilibrium steady state (NESS) is set up by having the plates at a different temperature, or chemical potential. In the former case the imposed temperature gradient will induce an average concentration gradient $\nabla\phi_0$ so that the average pressure gradient is zero. We assume that these average gradients are basically constant, or that there is a linear temperature and concentration profile so  that $\nabla\phi_0=\Delta\phi_0/L$, where $\Delta\phi_0$ is the concentration difference between the two plates. The concentration and velocity fields in the assumed NESS are then,
\begin{equation}
\phi_0(\mathbf{r})=\phi_{00}+\frac{\Delta\phi_0}{L}z,
\end{equation}
and
\begin{equation}
\mathbf{u}_0=0
\end{equation}
with $\phi_{00}$ a constant. Note that Eq.(2.5) implies a nonzero particle current,
\begin{equation}
J_z=-D\frac{\partial\phi_0}{\partial z}=-D\frac{\Delta\phi_0}{L}.
\end{equation}
This means there is a net flux of particles entering the system from the top plate and leaving the system from the bottom plate.

Finally, we further assume that we can ignore the dynamical temperature fluctuations since they decay on a relatively fast time scale \cite{Loi_et_al_2008, Marconi_et_al_2017}.

\section{THE INSTABILITY}
\label{sec:III}

Linearizing the Eqs.(2.1) and (2.2) about the NESS allows us to determine the stability of the solution. Assuming no-slip boundary conditions we use the Fourier representations,
\begin{equation}
\nonumber
\begin{split}
 (\delta\phi(\mathbf{r},t), u_z(\mathbf{r}, t))=\frac{2}{L}\sum_{n=1}\int\frac{d\omega}{2\pi}\int_{\mathbf{k}_{\perp}}\\e^{i\mathbf{k}_{\perp}\cdot\mathbf{r}_{\perp}-i\omega t}\sin(\frac{n\pi z}{L})(\delta\phi(\mathbf{k}, \omega), u_z(\mathbf{k}, \omega)). 
 \end{split}
 \end{equation}
Here $\mathbf{k}_{\perp}=(k_x, k_y)$, $\mathbf{k}=(\mathbf{k}_{\perp}, n\pi/L)$,  and $\mathbf{r}_{\perp}=(x,y)$.  

The interesting eigenfrequencies are derived next. From the basic physics of the instability problem we are interested in how shear modes in the velocity fluctuations couple to the diffusion modes in the swimmer density equation. This implies we can take the fluid to be incompressible and use the condition $\nabla\cdot\mathbf{u}=0$ on Eq.(2.2) to eliminate the pressure in terms of $\Sigma$ and $\mathbf P$. The Fourier transforms of the Eqs.(2.1) and (2.2) linearized about the NESS are
\begin{equation}
[-i\omega+Dk^2]\delta\phi(\mathbf{k},\omega)=-(\partial_z\phi_0)u_z(\mathbf{k},\omega)
\end{equation}
\begin{equation}
[-i\omega+\nu k^2]u_z(\mathbf{k},\omega)=\zeta k_{\perp}^2(\partial_z\phi_0)\delta\phi(\mathbf{k},\omega)+Q(\mathbf{k},\omega)
\end{equation}
with $Q$ a projected part of the noise term $\nabla\cdot\mathbf P$, $Q=ik_{\alpha}[P_{z\alpha}-\delta_{z\alpha}k_{\mu}k_{\beta}P_{\mu\beta}/k^2]$, that has the correlation,
\begin{equation}
\langle Q(\mathbf{k},\omega)Q(\mathbf{k}',\omega')\rangle=(2\pi)^{d}\frac{L}{2}\delta_{k_z,k_z'}\delta(\mathbf{k}_{\perp}+\mathbf{k}_{\perp}')\delta(\omega+\omega')2\nu k_{\perp}^2.
\end{equation}
Using these equations gives (here we use units where $k_BT=1$), 
\begin{equation}
\langle |\delta\phi(\mathbf{k},\omega)|^2\rangle=\frac{(\partial_z\phi_0)^22\nu k_{\perp}^2}{|F(\mathbf{k},\omega)|^2}
\end{equation}
with
\begin{equation}
F(\mathbf{k},\omega)=[-i\omega+Dk^2][-i\omega+\nu k^2]+\zeta k_{\perp}^2(\partial_z\phi_0)^2.
\end{equation}
Solving for the roots of $F(\mathbf{k}, \omega)$ gives,

\begin{equation}
\omega_{\pm}=\frac{-ik^2}{2}(D+\nu)\pm\frac{i}{2}\sqrt{k^4(D-\nu)^2-4\zeta(\partial_z\phi_0)^2k_{\perp}^2},
\end{equation}
with $k^2=k_{\perp}^2+k_z^2=k_{\perp}^2+n^2\pi^2/L^2$. 

For $\zeta=0$ Eq.(3.6) gives a shear mode and a diffusion mode. For $\zeta>0$ and large, the two modes change from being diffusive to propagating. More interestingly, for $\zeta<0$, and large magnitude, the $\omega_+$ mode becomes unstable. Structurally this is very similar to what happens at the Rayleigh-Bernard (RB) instability \cite{Kirkpatrick_Cohen_1983}. It first occurs at $k_z=k_{\perp}=\pi/L$. The analog of the Rayleigh number is $N\equiv(\Delta\phi_0)^2|\zeta|/D\nu$. Note that physically this is similar to the Rayleigh number, diffusion and viscosity suppress the instability (make $N$ smaller), while the activity parameter takes the place of gravity in driving the instability. The critical $N$ for the instability is $N_c=4\pi^2$. Near the instability,
\begin{equation}
\omega_+\approx -i\frac{2\nu D}{(\nu+D)}([k_{\perp}-\frac{\pi}{L}]^2+\frac{\pi^2}{L^2}\epsilon),
\end{equation}
where $N=N_c(1-\epsilon)$, with $\epsilon\ll1$. Note the implied critical slowing down as the instability is approached. The average flow pattern and concentration field for $N>N_c$ are discussed and illustrated in Sec.IV.

The feedback mechanism that causes the instability is that is that a convective force $\propto|\zeta|(\partial_z\phi_0)^2$ competes with dissipative forces $\propto D\nu$. For sufficiently large $|\zeta|(\partial_z\phi_0)^2$ the convective force wins and there is an instability. The linear mathematics of this instability are identical, for example, to the instability in the Richardson combat/arms race model discussed in \cite{Kibble_Berkshire_2004}. If we exclude complex roots, we see below that the non-linear mathematics of the instability are similar to the pitchfork bifurcation that occurs in the RB problem.

\section{THERMAL FLUCTUATION EFFECTS BELOW THE INSTABILITY}
\label{sec:IV}

It is interesting to compute various equal time correlation functions that characterize the generic long-ranged correlations in the NESS, and especially their amplification as the instability is approached. Of particular interest is the static structure factor, $S(k_{\perp},k_z)$, defined by, 

\begin{equation}
\langle |\delta\phi(\mathbf{k})|^2\rangle=S(k_{\perp},k_z)=\int\frac{d\omega}{2\pi}\langle |\delta\phi(\mathbf{k},\omega)|^2\rangle.
\end{equation}
Using Eqs.(3.4)-(3.6) gives,
\begin{equation}
S(k_{\perp},k_z)=\frac{(\partial_z\phi_0)^2k_{\perp}^2}{D(\nu+D)k^2}\frac{1}{[k^4-Nk_{\perp}^2/L^2]}
\end{equation}
(for $\zeta>0$ change the sign of $N$). Note that is the absence of activity ($N=0$), this long-range correlation is analogous to the experimentally well verified \cite{Law_et_al_1990} one that appears in a simple fluid
in a temperature gradient \cite{Kirkpatrick_Cohen_Dorfman_1982A}. In the presence of activity it is enhanced (suppressed) compared to the passive fluid result for $\zeta<0$ ($\zeta>0$). Near the instability the singular contribution is,
\begin{equation}
S(k_{\perp},k_z=\pi/L)_{\mathrm{sing}}\approx\frac{(\Delta\phi_0)^2}{8\pi^2D(\nu+D)}\frac{1}{[(k_{\perp}-\pi/L)^2+\pi^2\epsilon/L^2]}.
\end{equation}

The NE Casimir pressure, $p_{NE}(L)$,  or force is also of interest. For a passive fluid in a NESS it has been discussed in great detail elsewhere \cite{Kirkpatrick_DeZarate_Sengers_2013, Kirkpatrick_DeZarate_Sengers_2014, Kirkpatrick_DeZarate_Sengers_2016a, Kirkpatrick_DeZarate_Sengers_2016b, Aminov_et_al_2015}. Physically, non-linear long-range fluctuations renormalize the pressure \cite{Kardar_Golestanain_1999},
\begin{equation}
p_{NE}(L)=\frac{1}{2}\Big(\frac{\partial^2p}{\partial\phi^2}\Big)_T\overline{\langle\delta\phi(\mathbf{r})^2\rangle},
\end{equation}
where the over-line denotes a spatial average.
For small $|N|$ the equal time correlation function in Eq.(4.4) is,
\begin{equation}
\overline{\langle\delta\phi(\mathbf{r})^2\rangle}_{|N|\ll 1}\approx\frac{(\Delta\phi_0)^2}{48\pi LD(D+\nu)}
\end{equation}
Near the instability there is a singular contribution given by,
\begin{equation}
\overline{\langle\delta\phi(\mathbf{r})^2\rangle}_{N=N_c(1-\epsilon)}\approx\frac{(\Delta\phi_0)^2}{32\pi^2LD(\nu+D)\sqrt{\epsilon}}.
\end{equation}
For $\zeta>0$ and $|N|\gg 1$ one obtains,
\begin{equation}
\overline{\langle\delta\phi(\mathbf{r})^2\rangle}_{|N|\gg 1}\approx\frac{(\Delta\phi_0)^2}{16LD(\nu+D)\sqrt{|N|}}.
\end{equation}
Note that these Casimir forces are huge at long distances compared to the vacuum fluctuation or critical Casimir forces that decay as $1/L^4$ and $1/L^3$, respectively, for large $L$.  This $L$-dependence  can be used to distinguish this force from other, short ranged, forces. Also, we again see that positive activity suppresses fluctuations effects, while negative activity enhances fluctuations.

As the instability is approached, the transport coefficients themselves become singularly renormalized. This leads to a sort of Ginzburg criterion for ignoring nonlinear fluctuation effects as the instability is approached: If the renormalization effects are small, then the nonlinear fluctuations can be ignored and the instability analysis given above is valid. To illustrate this, the mode-coupling renormalization of the diffusion coefficient, $\delta D$ , can be determined as follows. The Eq.(2.1) defines a concentration current in the $z$-direction given by,
\begin{equation}
J_z=-u_z(\mathbf{r})\delta\phi(\mathbf{r})+D\partial_z\phi(\mathbf{r})
\end{equation}
Averaging over the fluctuations gives an equation for $\delta D$,
\begin{equation}
\delta D=-\frac{\langle u_z(\mathbf{r})\delta\phi(\mathbf{r})\rangle}{(\partial_z\phi_0)}.
\end{equation}
Near the instability the singular contribution can be determined using Eqs.(3.1)-(3.3) and Eq.(3.7),
\begin{equation}
\delta D\approx\frac{1}{16L(D+\nu)\sqrt{\epsilon}}.
\end{equation}
Numerically this is a very small (it is a $1/L$ effect) perturbation on the bare $D$ unless one is extraordinarily close to the instability. A similar result is obtained for the thermal diffusion coefficient near the RB instability \cite{Kirkpatrick_Cohen_1983}. In practice this means that more sophisticated self-consistent or renormalization group-like treatments are not needed to describe these instabilities \cite{Swift_Hohenberg_1977}.

\section{ACTIVE FLUID BEHAVIOR ABOVE THE INSTABILITY}
\label{sec:V}

To complete the description of the active fluid  hydrodynamic instability  we consider the active fluid average NE motion above the instability threshold by constructing a three-mode Lorenz-like model \cite{Lorenz_1993}. Only for $N>N_c$ does it become clear that the  fluid behavior at this hydrodynamic instability is qualitatively different than the passive fluid behavior near a Rayleigh-Benard instability. The spatial structure of the convection rolls that occur for $N>N_c$ are determined by the critical wavenumbers being $k_{\perp}=k_z=\pi$ (here we use units where $L=1$) and by taking the fluid to be incompressible ($\nabla\cdot\mathbf{u}=0$). We are interested here in the average fluid motion, so that the Langevin force in Eq.(2) can be neglected. Taking the curl of Eq.(2.2) eliminates the pressure so that equations of motion, Eq.(2.1) and Eq.(2.22), are for the wall normal and wall transverse velocities $(u_z, u_x)$, and $\delta\phi=\phi-\phi_0$. Consistent with this we define,
\begin{equation}
u_z=A(t)\cos\pi x\sin\pi z \nonumber
\end{equation}
\begin{equation}
u_x=-A(t)\sin\pi x\cos\pi z
\end{equation}
\begin{equation}
\delta\phi=B(t)\cos\pi x\sin\pi z+C(t)\sin 2\pi z \nonumber
\end{equation}
Inserting the representations of $(u_z, u_x,\delta\phi)$  into these equation of motion gives a set of coupled nonlinear ordinary differential equations for $(A,B,C)$ with multiplicative trigonometric functions. Multiplying the $(\dot A,\dot B,\dot C)$ equations by $(\int dz\sin(\pi z), \int dz\sin(\pi z),\int dz\sin(2\pi z))$, respectively, leads to the equations,
\begin{equation}
\dot A=-2\pi^2\nu A-|\zeta|(\partial_z\phi_0)\pi^2B-|\zeta|\pi^3BC
\end{equation}
\begin{equation}
\dot B=-2\pi^2 DB-(\partial_z\phi_0)A+\pi AC
\end{equation}
\begin{equation}
\dot C=-4\pi^2DC-\frac{\pi}{2}AB.
\end{equation}
Defining a dimensionless time by $\tau=2\pi^2Dt$ and $X=A/2^{3/2}D\pi$, $Y=\pi B/\sqrt{2}(\partial_z\phi_0)$ and $Z=-\pi C/(\partial_z\phi_0)$ leads to the scaled equations of motion,
\begin{equation}
\dot X=\sigma(-X+rY+rYZ) \nonumber
\end{equation}
\begin{equation}
\dot Y=-XZ+X-Y
\end{equation}
\begin{equation}
\dot Z=-2Z+XY \nonumber
\end{equation}
These are the generalized Lorenz equations for the active matter instability.
In these equations $r=N/N_c$ and $\sigma=\nu/D$ is the Prandtl number for this system \footnote{This Prandtl number will generally be very large compared to the Prandtl number in the RB problem because particle diffusion coefficients are typically very small compared to thermal diffusion coefficients.}. The non-linear term in the $\dot X$ equation reflects the activity non-linearity and is not present in the Rayleigh-Bernard problem. 

The fixed points of these Lorenz equations are,
\begin{equation}
X=Y=Z=0 \quad(\mathrm{stable\quad for\quad} r<1)
\end{equation}
and for $r>1$,
\begin{equation}
\frac{X^2}{2}=r-1\pm\sqrt{(r-1)^2+r-1}\quad (\mathrm{two\quad\ real\quad roots})\nonumber
\end{equation}
\begin{equation}
Y=\frac{X}{(1+\frac{X^2}{2})}
\end{equation}
\begin{equation}
Z=\frac{XY}{2}\nonumber.
\end{equation}
The flow pattern for $N>N_c$ is shown in Fig.1a, while the concentration profile is shown in Fig.1b.
\begin{figure}[h]
	\includegraphics[scale=0.15]{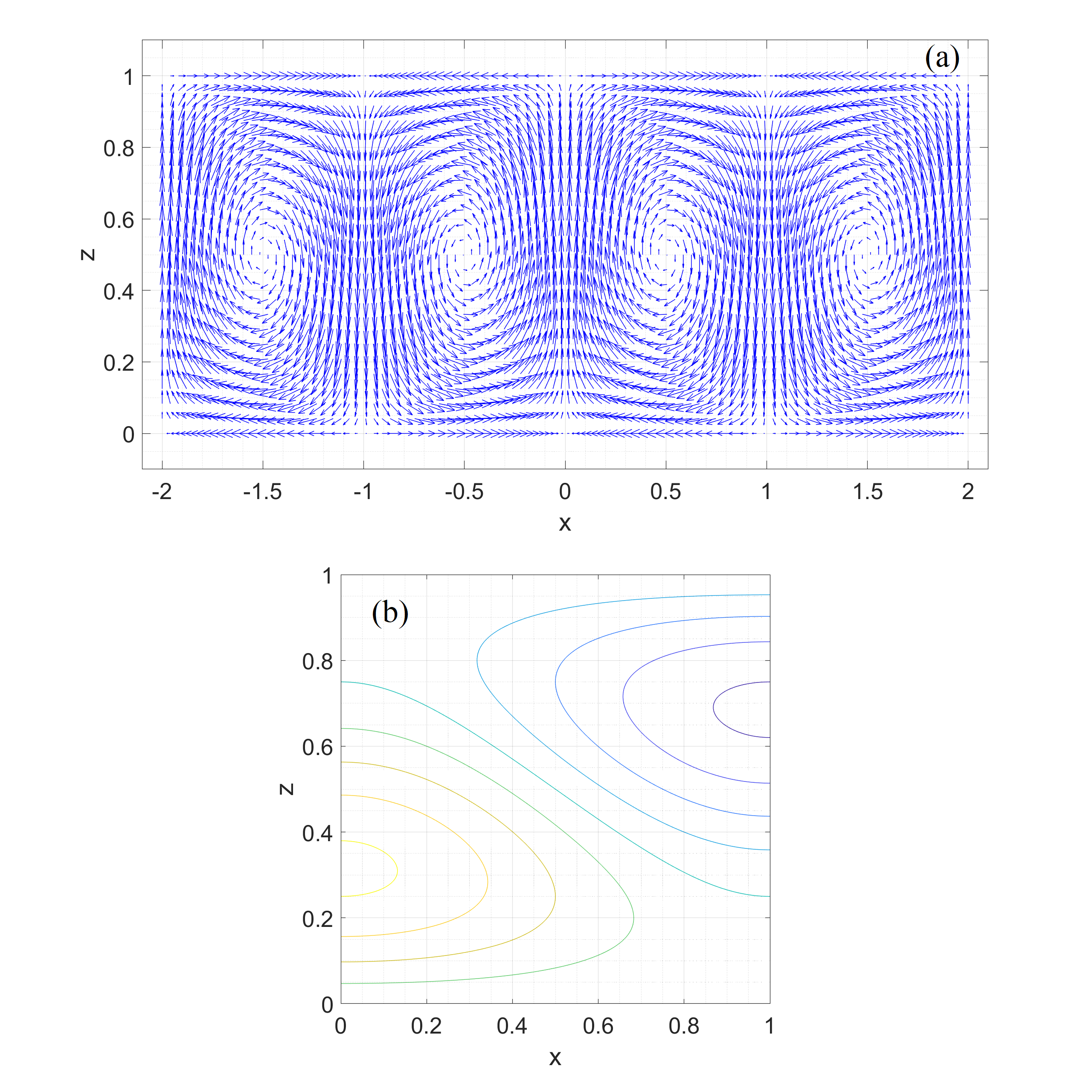}
	\caption{(a) The cellular flow pattern after the onset of the instability using the generalized Lorenz equations. (b) Typical concentration contours in a single cell.}
\end{figure}

Compared to the RB Lorenz equations \cite{Kibble_Berkshire_2004, Ott_1993} there are at least three interesting features associated with these Lorenz equations that warrant further study, i) The physical fixed points close to and above the convective instability are $(X,Y,Z)=(\pm 2^{1/2}(r-1)^{1/4}, \pm 2^{1/2}(r-1)^{1/4}, (r-1)^{1/2})$. In the RB problem, $(r-1)^{1/4}$ is replaced  with $(r-1)^{1/2}$. The convective transport is proportional to $Z=\sqrt{r-1}$, which is non-analytic in the control parameter $r-1$, unlike in the RB problem \cite{Lorenz_1993,Ott_1993} where $Z=r-1$ and is therefore analytic. Further, unlike the RB problem for $r>>1$, the transport saturates in this case. The difference between the two transports is shown in Fig.2. ii) The crucial nonlinearity in the $\dot X$ equation is proportional to $r$, and thus increases with driving causing the time independent solution given by Eqs.(5.7) to become unstable at a smaller $r$ than in the RB problem. This implies that the Lorenz equations for this system are a more realistic representation of the active fluid hydrodynamics than the RB Lorenz equations are for the passive fluid hydrodynamics. iii) There are two additional complex fixed points of the Eqs.(5.7), compared to the RB problem. This suggests the Hopf-bifurcation and the transition to turbulence in this system will be qualitatively different than in the RB problem.
\begin{figure}[h]
	\includegraphics[scale=0.15]{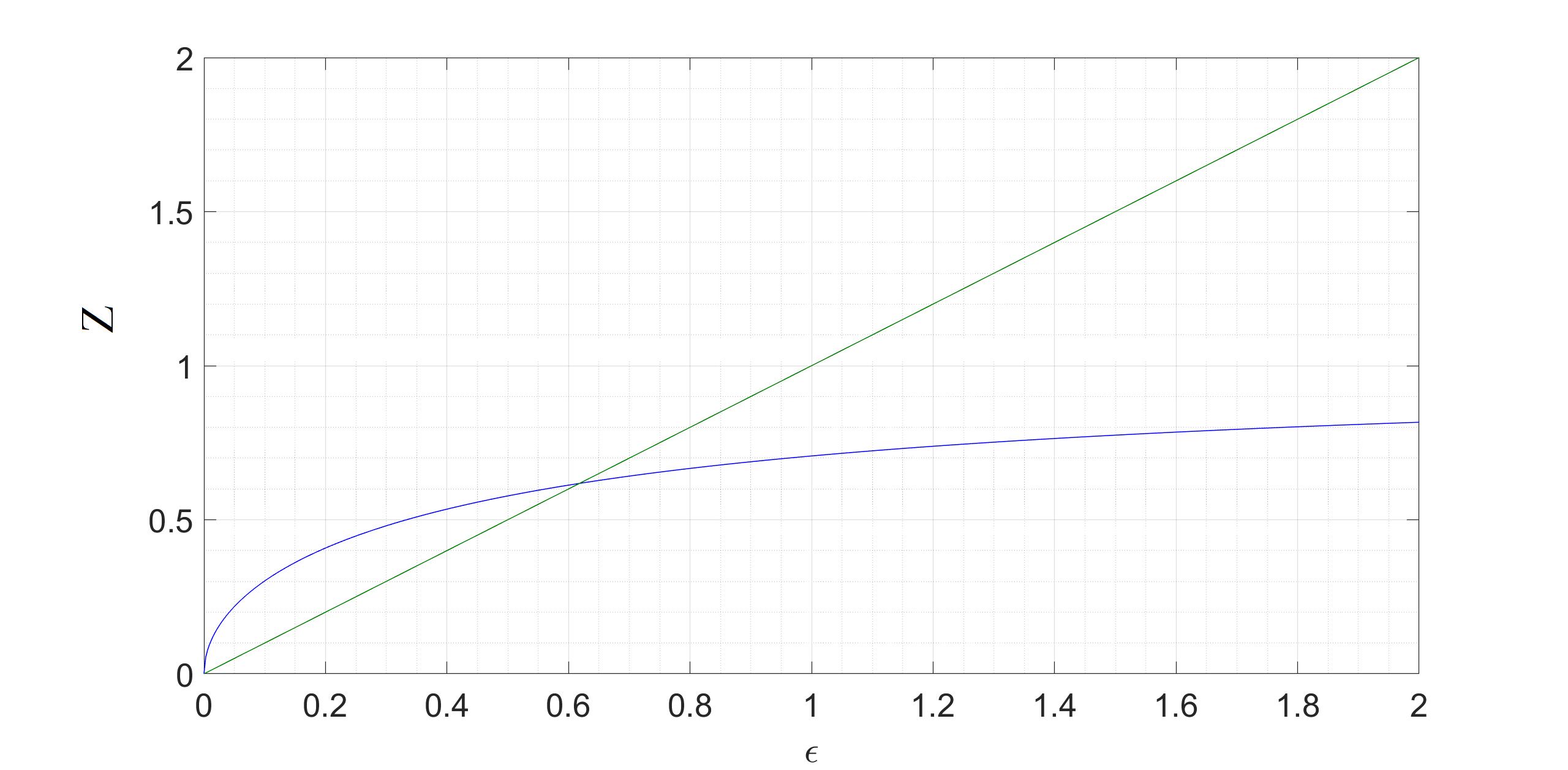}
	\caption{The concentration flux from one plate to another as a function of the parameter $\epsilon=r-1$. The saturation at high values and the non-analytic behavior near the threshold is a distinctive feature of the present model. In contrast, the standard Lorenz model shows a linear dependence.}
\end{figure}

The steady state solution given by Eqs.(5.7) physically means that the average fluid motion for $r>1$ consist of time-independent convection rolls in addition to the linear $\phi_0$ profile. This roll solution will become unstable at a Hopf-bifurcation, $r_H$, to time-dependent motion. The instability point can be determined by inserting the Eqs.(5.7) into the Eqs.(5.5) and linearizing about the steady state. Since the Eqs.(5.5) are three coupled equations, this defines a cubic eigenvalue problem. Assuming a $e^{i\omega t}$ time dependence, with $\omega=\omega_R+i\omega_I$, implies that at the Hopf-bifurcation $\omega_I=0$. That is, linear perturbations about the steady-state solution no longer decay in time at $r=r_H$. At $r_H$ the complex cubic equation for $\omega_R$ has a real and imaginary part, both of which must be satisfied. These two equations are,
\begin{equation}
\omega_R^2=2+X^2+3\sigma-\sigma r(1-Z^2)-\sigma rY^2
\end{equation}
and
\begin{equation}
(3+\sigma)\omega_R^2=\sigma(2+X^2)+2\sigma r(3Z^2-1)-\sigma rY^2,
\end{equation}
with $(X,Y,Z)$ given by the Eqs.(5.7).
Equating $\omega_R^2$ from these two equations gives $r_H(\sigma)$. As noted already, $\sigma$ will typically be quite large. From the two equations above it follows that $r_H$ grows with $\sigma$. For example, in the very large sigma limit it is easily shown that $r_H\approx 3\sigma/4$. To do better, we use that the Eqs.(5.7) for large $r$ give,  
\begin{equation}
X^2=4\epsilon+1+O(\epsilon^{-1}) \nonumber
\end{equation}
\begin{equation}
Y^2=\frac{1}{\epsilon}(1-\frac{5}{4\epsilon}+O(\epsilon^{-2}))
\end{equation}
\begin{equation}
Z^2=1-\frac{1}{\epsilon}+\frac{1}{\epsilon^2}+O(\epsilon^{-3}) \nonumber
\end{equation}
with $\epsilon=r-1$. Using these results in equating the two equations and retaining leading order and next to leading order terms for large $\sigma$  leads to a quadratic equation for $r_H$ that when solved gives,
\begin{equation}
r_H=1+\frac{\sigma^2+6\sigma+\sigma\sqrt{\sigma^2+16\sigma+24}}{8(\sigma-3)}
\end{equation}
For large $\sigma$ the frequency at the critical point is $\omega\approx\pm\sqrt{8\sigma(r_H-1)}$. We note that at $\sigma=10$, an exact solution of the cubic equation for the Hopf-bifurcation gives $r_H\approx 7.19$ \footnote{Eq. 18 gives $r_H\approx 6.87$ for $\sigma=10$. For $\sigma=50$, both the full cubic equation and Eq.(18) give $r_H\approx16.1$.}. In contrast, for the RB Lorenz equations there is a Hopf-bifurcation at $r_H=\sigma(\sigma+5)/(\sigma-3)$, so that at $\sigma=10$, this $r_H\approx 21.4$. These two very different values of $r_H$ are consistent with the notion that the Lorenz equations given by Eq.(5.5) are a better model for the active matter full hydrodynamic equations than the original Lorenz equations are for the RB problem. Physically this is plausible because both sets of Lorenz equations ignore the fluid velocity convective nonlinearity in the $\dot X$ equation, but in the active matter case, this nonlinearity will be sub-leading to the activity nonlinearity if the activity coefficient is large.

In Fig.3 the active fluid behavior in the $X-Y$ plane is illustrated. In particular we show that below the Hopf-point (for $\sigma=10$) there is a very clear limit cycle. Close to the Hopf-point the basin of attraction of the stable fixed point, Eq.(5.7), is extremely small: One needs initial conditions very close to the fixed point to reach it, otherwise the flow is to the limit cycle. The separatrix between the fixed point and the limit cycle is clear. There is another separatrix between the limit cycle and a strange attractor. In the passive fluid Lorenz model there is only a coexisting limit cycle and chaos, i.e., there is never a clear cut limit cycle. In general it appears that in comparison to the RB problem, the smaller values of $(X, Y, Z)$ at the fixed point, Eq.(5.7), seem to make the active fluid more stable near $r_H$.
\begin{figure}[h]
	\includegraphics[scale=0.15]{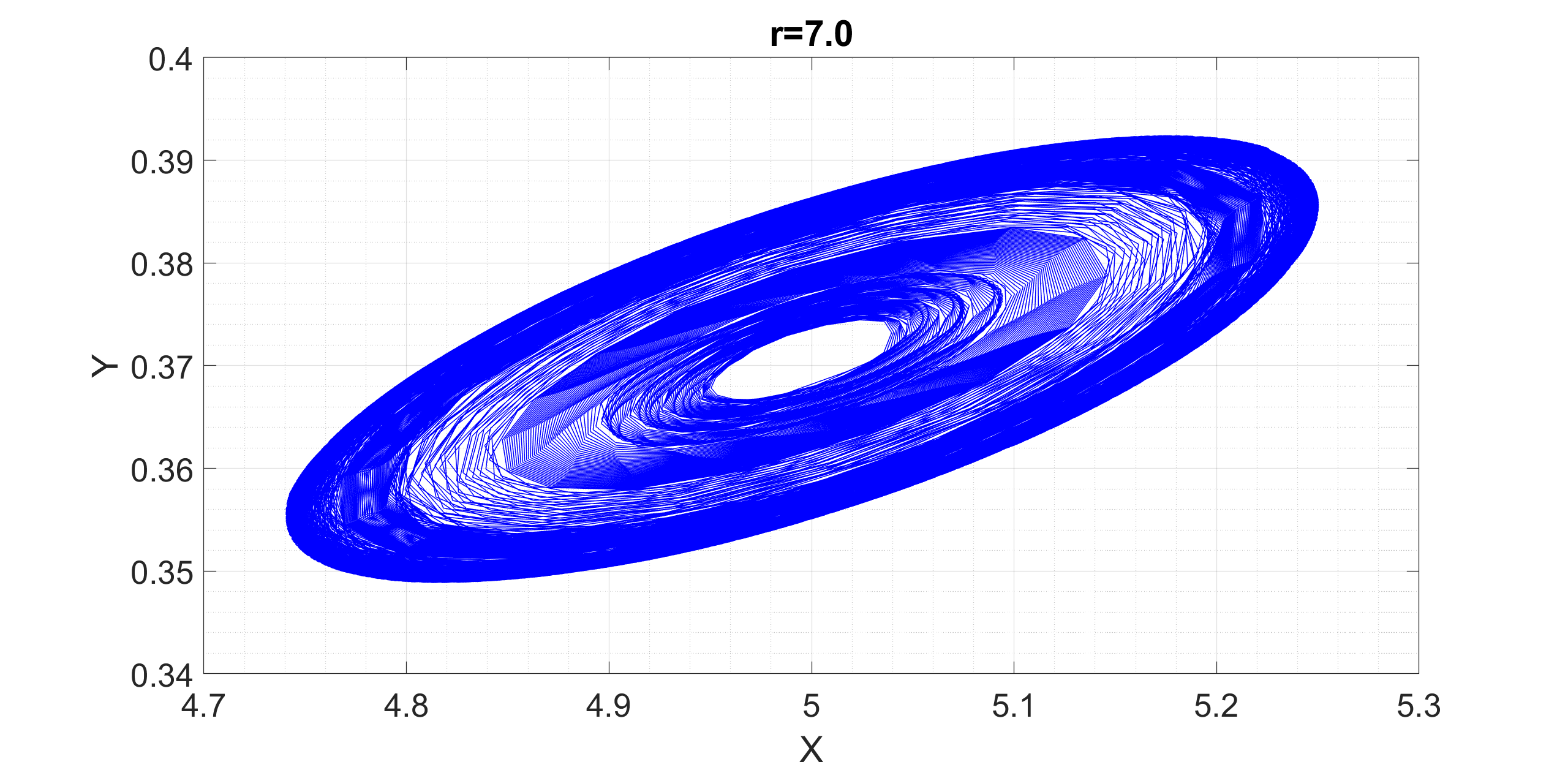}
	\caption{The hysteretic limit cycle just below the threshold of the Hopf-bifurcation. Inside the unshaded elliptic region of the figure there is a closed curve which is the locus of initial conditions which separate the basin of attraction of the stable fixed point from the basin of the limit cycle. }
\end{figure}

\section{DISCUSSION}
\label{sec:V}

Driven active matter has been shown to exhibit a new type of convective instability reminiscent of the RB instability in passive fluids. As the instability is approached from below, $N\rightarrow N_c^-$, various singular correlation functions have been computed. For $N>N_c$ the average fluid behavior has been characterized, and a Hopf-bifurcation to time dependent flow has been illustrated. 

We conclude with a number of further remarks:
\begin{enumerate}
\item The presumed size of $\zeta$ here and in \cite{Tiribocchi_et_al_2015} is quite large. In simple passive fluids at liquid state densities we can estimate the scale of $\zeta$ as follows. If $\phi$ is dimensionless then $\zeta$ has the dimensions of $\ell^4/\tau^2=v^2\ell^2$. Here $\ell$ is a length, $\tau$ is a time, and $v$ is a velocity. $\ell$ is the larger of the molecular diameter, $\sigma$,  and the mean-free-path, which for liquid state densities would be $\sigma$, and $v$ is the thermal velocity. For water at standard temperature and pressure, this would numerically give $|\zeta|\approx 2\cdot 10^{-6}\mathrm{cm^4/sec^2}$, which is about an order of magnitude larger than $D\nu$ in water. The value of $N_c$ and this suggest that the activity part of $|\zeta|$ plays a qualitatively new role when it is larger than the passive one by a factor of $100$ to $1000$.
\item To within factors the condition for the instability is $|\zeta|(\Delta\phi_0)^2\propto D\nu$. We can compare this condition to, for example, the condition for the bending instability of extensile filaments in active matter \cite{Ramaswamy_2010, Simha_Ramaswamy_2002}. First we note that because of the gradients in Eq.(2.4), $|\zeta|/L^2$ corresponds to $Wc_0$ in \cite{Ramaswamy_2010}. Using Stokes law, $D\propto k_BT/a\eta$, witn $a$ a mesoscopic length, we can write the our instability condition as $Wc_0\propto k_BT/aL^2(\Delta\phi_0)^2$. Apart from the factor $(\Delta\phi_0)^2$, expected for  a gradient induced instability, this is basically the condition for the bending instability \cite{Ramaswamy_2010}. We conclude that for moderate $\Delta\phi_0$ the NE instability discussed here could be experimentally relevant.
\item Experimental studies of (wet) active matter systems with an average temperature (or chemical potential) gradient would be very interesting. If the activity coefficient is large enough, the hydrodynamic instability discussed here will be apparent. It could be distinguished from the usual RB instability because it will occur when the fluid is heated both from below and above. Further, the generalized Lorenz equations indicate that the fluid behavior for $N>N_c$ will be very different than what occurs in a passive system convective instability.
\item In general in both passive and active fluids a gradient expansion breaks down after Naiver-Stokes order (in simple fluids in three-dimensions) and the generalized description must be non-local \cite{Ernst_Dorfman_1975, Ernst_et_al_1978, Dorfman_Kirkpatrick_Sengers_1994, Belitz_Kirkpatrick_Vojta_2005}. Technically one finds divergences in the calculation of higher order or Burnett transport coefficients such as $\zeta$. These singular renormalization will have a scale set by the small passive generalized transport coefficients and will presumably not be important. This deserves further study.

\end{enumerate}

\medskip
Discussions with Jan Sengers are gratefully acknowledged. JKB would like to thank the IPST at the University of Maryland for support during the initial stages of this work. In addition, this work was supported by the National Science Foundation under Grant
No. DMR-1401449.

\end{document}